\documentclass[3p]{elsarticle}

\usepackage[utf8]{inputenc}
\usepackage[T1]{fontenc}
\usepackage{uniinput}
\usepackage{graphicx}

\usepackage{hyperref}
\usepackage{amsmath,amssymb,amsfonts,amsthm,bbm,dsfont,slashed,xfrac,nicefrac}
\usepackage{mathrsfs}
\usepackage{axodraw2}
\usepackage{fancyvrb}

\newcommand{\lagrangian}{\mathscr{L}}
\newcommand{\del}{\partial}
\newcommand{\f}[1]{f^{\ensuremath{#1}}}
\newcommand{\cbar}{\overline{\mbox{$c$}}}
\newcommand{\psibar}{\overline{\mbox{$ψ$}}}

\newcommand{\FGqqProp}[1]{
\hspace{-4pt}
 \raisebox{-.5\height}{
  \SetScale{#1}
  \begin{axopicture}(70,20)(0,40)
    \Line[arrow](60,50)(10,50)
  \end{axopicture}
 }
}
\newcommand{\FGgluProp}[1]{
\hspace{-4pt}
 \raisebox{-.5\height}{
  \SetScale{#1}
  \begin{axopicture}(70,20)(0,40)
    \Photon(10,50)(60,50){3.5}{5}
  \end{axopicture}
 }
}
\newcommand{\FGhhProp}[1]{
\hspace{-4pt}
 \raisebox{-.5\height}{
  \SetScale{#1}
  \begin{axopicture}(70,20)(0,40)
    \Line[arrow,dash,dsize=4](60,50)(10,50)
  \end{axopicture}
 }
}
\newcommand{\FGhhGluVertex}[1]{
\hspace{-4pt}
 \raisebox{-.5\height}{
  \SetScale{#1}
  \begin{axopicture}(80,100)(10,0)
    \Photon(50,50)(50,85){4}{3}
    \Line[arrow,dash,dsize=4](80,32)(50,50)
    \Line[arrow,dash,dsize=4](50,50)(20,32)
  \end{axopicture}
 }
}
\newcommand{\FGqqGluVertex}[1]{
\hspace{-4pt}
 \raisebox{-.5\height}{
  \SetScale{#1}
  \begin{axopicture}(80,100)(10,0)
    \Photon(50,50)(50,85){4}{3}
    \Line[arrow](80,32)(50,50)
    \Line[arrow](50,50)(20,32)
  \end{axopicture}
 }
}
\newcommand{\FGtriGluVertex}[1]{
\hspace{-4pt}
 \raisebox{-.5\height}{
  \SetScale{#1}
  \begin{axopicture}(80,100)(10,0)
    \Photon(50,50)(50,85){4}{3}
    \Photon(50,50)(80,32){4}{3}
    \Photon(50,50)(20,32){4}{3}
  \end{axopicture}
 }
}
\newcommand{\FGtetraGluVertex}[1]{
\hspace{-4pt}
 \raisebox{-.5\height}{
  \SetScale{#1}
  \begin{axopicture}(80,100)(10,0)
    \Photon(50,50)(75,75){4}{3}
    \Photon(50,50)(75,25){4}{3}
    \Photon(50,50)(25,25){4}{3}
    \Photon(50,50)(25,75){4}{3}
  \end{axopicture}
 }
}
\newcommand{\FGggProp}[1]{
\hspace{-4pt}
 \raisebox{-.5\height}{
  \SetScale{#1}
  \begin{axopicture}(70,20)(0,40)
    \DoublePhoton(10,50)(60,50){3.5}{5}{1.5}
  \end{axopicture}
 }
}
\newcommand{\FGhhgVertex}[1]{
\hspace{-4pt}
 \raisebox{-.5\height}{
  \SetScale{#1}
  \begin{axopicture}(80,100)(10,0)
    \DoublePhoton(50,50)(50,85){4}{3}{1.5}
    \Line[arrow,dash,dsize=4](80,32)(50,50)
    \Line[arrow,dash,dsize=4](50,50)(20,32)
  \end{axopicture}
 }
}
\newcommand{\FGgggVertex}[1]{
\hspace{-4pt}
 \raisebox{-.5\height}{
  \SetScale{#1}
  \begin{axopicture}(80,100)(10,0)
    \DoublePhoton(50,50)(50,85){4}{3}{1.5}
    \DoublePhoton(50,50)(80,32){4}{3}{1.5}
    \DoublePhoton(50,50)(20,32){4}{3}{1.5}
  \end{axopicture}
 }
}
\newcommand{\FGggggVertex}[1]{
\hspace{-4pt}
 \raisebox{-.5\height}{
  \SetScale{#1}
  \begin{axopicture}(80,100)(10,0)
    \DoublePhoton(50,50)(75,75){4}{3}{1.5}
    \DoublePhoton(50,50)(75,25){4}{3}{1.5}
    \DoublePhoton(50,50)(25,25){4}{3}{1.5}
    \DoublePhoton(50,50)(25,75){4}{3}{1.5}
  \end{axopicture}
 }
}
\newcommand{\FGgggggVertex}[1]{
\hspace{-4pt}
 \raisebox{-.5\height}{
  \SetScale{#1}
  \begin{axopicture}(80,100)(10,0)
    \DoublePhoton(50,50)(50,85){4}{3}{1.5}
    \DoublePhoton(50,50)(83,61){4}{3}{1.5}
    \DoublePhoton(50,50)(70.5,22){4}{3}{1.5}
    \DoublePhoton(50,50)(29.5,22){4}{3}{1.5}
    \DoublePhoton(50,50)(17,61){4}{3}{1.5}
  \end{axopicture}
 }
}
\newcommand{\FGggggggVertex}[1]{
\hspace{-4pt}
 \raisebox{-.5\height}{
  \SetScale{#1}
  \begin{axopicture}(80,100)(10,0)
    \DoublePhoton(50,50)(50,85){4}{3}{1.5}
    \DoublePhoton(50,50)(82,67.5){4}{3}{1.5}
    \DoublePhoton(50,50)(82,32.5){4}{3}{1.5}
    \DoublePhoton(50,50)(50,15){4}{3}{1.5}
    \DoublePhoton(50,50)(18,32.5){4}{3}{1.5}
    \DoublePhoton(50,50)(18,67.5){4}{3}{1.5}
  \end{axopicture}
 }
}
\newcommand{\FGtriGluContrBlob}[1]{
\hspace{-4pt}
 \raisebox{-.5\height}{
  \SetScale{#1}
  \begin{axopicture}(70,70)(15,15)
    \Photon[width=.34](50,50)(50,85){3}{5}
    \Line[arrow,arrowpos=0.7,arrowlength=7.5,arrowwidth=5,arrowinset=0](50,95)(50,85)
    \Photon[width=.34](50,50)(80,32){3}{5}
    \Photon[width=.34](50,50)(20,32){3}{5}
    \GCirc(50,50){10}{0.82}
    \end{axopicture}
 }
}
\newcommand{\FGtriGluTransLBlob}[1]{
\hspace{-4pt}
 \raisebox{-.5\height}{
  \SetScale{#1}
  \begin{axopicture}(70,70)(15,15)
    \Line[width=.34,dash,dsize=4,arrow](50,50)(50,85)
    \Photon[width=.34](50,50)(80,32){3}{5}
    \Line[width=.34](50,50)(26,36)
    \GCirc(50,50){10}{0.82}
    \Text(20,32){\tiny T}
  \end{axopicture}
 }
}
\newcommand{\FGtriGluTransRBlob}[1]{
\hspace{-4pt}
 \raisebox{-.5\height}{
  \SetScale{#1}
  \begin{axopicture}(70,70)(15,15)
    \Line[width=.34,dash,dsize=4,arrow](50,50)(50,85)
    \Line[width=.34](50,50)(74,36)
    \Photon[width=.34](50,50)(20,32){3}{5}
    \GCirc(50,50){10}{0.82}
    \Text(80,32){\tiny T}
  \end{axopicture}
 }
}
\newcommand{\FGtriGluPropLBlob}[1]{
\hspace{-4pt}
 \raisebox{-.5\height}{
  \SetScale{#1}
  \begin{axopicture}(70,70)(15,15)
    \Line[width=.34,dash,dsize=4,arrow](50,50)(50,85)
    \Photon[width=.34](50,50)(80,32){3}{5}
    \Photon[width=.34](20,32)(40,44){3}{3.3}
    \Line[width=.34](50,50)(40,44)
    \Line[width=1](37,49)(43,39)
    \GCirc(30,38){5}{0.82}
  \end{axopicture}
 }
}
\newcommand{\FGtriGluPropRBlob}[1]{
\hspace{-4pt}
 \raisebox{-.5\height}{
  \SetScale{#1}
  \begin{axopicture}(70,70)(15,15)
    \Line[width=.34,dash,dsize=4,arrow](50,50)(50,85)
    \Photon[width=.34](80,32)(60,44){3}{3.3}
    \Line[width=.34](50,50)(60,44)    
    \Line[width=1](57,39)(63,49)
    \Photon[width=.34](50,50)(20,32){3}{5}
    \GCirc(70,38){5}{0.82}
  \end{axopicture}
 }
}
\newcommand{\FGtriGluGhBlob}[1]{
\hspace{-4pt}
 \raisebox{-.5\height}{
  \SetScale{#1}
  \begin{axopicture}(70,70)(15,15)
    \Line[width=.34,dash,dsize=4,arrow](50,75)(50,85)
    \Line[width=.34](50,60)(50,70)
    \Photon[width=.34](50,50)(50,60){3}{2}
    \Photon[width=.34](50,50)(80,32){3}{5}
    \Photon[width=.34](50,50)(20,32){3}{5}
    \Line[width=1](45,60)(55,60)
    \GCirc(50,70){5}{0.82}
    \end{axopicture}
 }
}

\newcommand{\FGgluPropContrCan}[1]{
\hspace{-4pt}
 \raisebox{-.5\height}{
  \SetScale{#1}
  \begin{axopicture}(90,20)(0,40)
    \Photon(20,50)(80,50){4}{5}
    \Line[arrow,arrowpos=0.6,arrowlength=11.25,arrowwidth=7.5,arrowinset=0](10,50)(20,50)
  \end{axopicture}
 }
}
\newcommand{\FGghPropContrCan}[1]{
\hspace{-4pt}
 \raisebox{-.5\height}{
  \SetScale{#1}
  \begin{axopicture}(90,20)(0,40)
    \Line[arrow,dash,dsize=6,arrowscale=1.5](70,50)(10,50)
    \Photon(80,50)(90,50){4}{1}
\Line[arrow,arrowpos=0.5,arrowlength=11.25,arrowwidth=7.5,arrowinset=0](70,50)(80,50)
  \end{axopicture}
 }
}
\newcommand{\FGtriGluContrCan}[1]{
\hspace{-4pt}
 \raisebox{-.5\height}{
  \SetScale{#1}
  \begin{axopicture}(70,70)(15,15)
    \Photon[width=.34](50,50)(50,85){3}{5}
    \Line[arrow,arrowpos=0.7,arrowlength=7.5,arrowwidth=5,arrowinset=0](50,95)(50,85)
    \Photon[width=.34](50,50)(80,32){3}{5}
    \Photon[width=.34](50,50)(20,32){3}{5}
    \end{axopicture}
 }
}
\newcommand{\FGauxiVertexCanL}[1]{
\hspace{-4pt}
 \raisebox{-.5\height}{
  \SetScale{#1}
  \begin{axopicture}(70,70)(15,15)
    \Line[width=.34,dash,dsize=4,arrow](50,50)(50,85)
    \Photon[width=.34](50,50)(80,32){3}{5}
    \Photon[width=.34](20,32)(35,41){3}{2.5}
    \Line[width=.34](50,50)(35,41)
    \Line[width=1](32,46)(38,36)
  \end{axopicture}
 }
}
\newcommand{\FGghVertexCanL}[1]{
\hspace{-4pt}
 \raisebox{-.5\height}{
  \SetScale{#1}
  \begin{axopicture}(70,70)(15,15)
    \Line[width=.34,dash,dsize=4,arrow](50,50)(50,85)
    \Line[width=.34,dash,dsize=4,arrow](20,32)(50,50)
    \Photon[width=.34](50,50)(80,32){3}{5}
    \Photon[width=.34](14,28)(20,32){3}{1}
\Line[width=.34,arrow,arrowpos=0.1,arrowlength=7.5,arrowwidth=5,arrowinset=0](23,34)(20,32)
  \end{axopicture}
 }
}
\newcommand{\FGauxiVertexCanR}[1]{
\hspace{-4pt}
 \raisebox{-.5\height}{
  \SetScale{#1}
  \begin{axopicture}(70,70)(15,15)
    \Line[width=.34,dash,dsize=4,arrow](50,50)(50,85)
    \Photon[width=.34](80,32)(65,41){3}{2.5}
    \Line[width=.34](50,50)(65,41)
    \Line[width=1](62,36)(68,46)
    \Photon[width=.34](50,50)(20,32){3}{5}
  \end{axopicture}
 }
}
\newcommand{\FGghVertexCanR}[1]{
\hspace{-4pt}
 \raisebox{-.5\height}{
  \SetScale{#1}
  \begin{axopicture}(70,85)(15,15)
    \Line[width=.34,dash,dsize=4,arrow](50,50)(50,85)
    \Line[width=.34,dash,dsize=4,arrow](80,32)(50,50)
    \Photon[width=.34](86,28)(80,32){3}{1}
\Line[width=.34,arrow,arrowpos=0.1,arrowlength=7.5,arrowwidth=5,arrowinset=0](77,34)(80,32)
    \Photon[width=.34](50,50)(20,32){3}{5}
  \end{axopicture}
 }
}
\newcommand{\FGfourGluContrCan}[1]{
\hspace{-4pt}
 \raisebox{-.5\height}{
  \SetScale{#1}
  \begin{axopicture}(70,100)(15,0)
    \Photon[width=.34](50,50)(75,75){3}{5}
    \Photon[width=.34](50,50)(75,25){3}{5}
    \Photon[width=.34](50,50)(25,25){3}{5}
    \Photon[width=.34](50,50)(25,75){3}{5}
    \Line[arrow,arrowpos=0.7,arrowlength=7.5,arrowwidth=5,arrowinset=0](18,82)(25,75)
  \end{axopicture}
 }
}
\newcommand{\FGfourGluAuxCanI}[1]{
\hspace{-4pt}
 \raisebox{-.5\height}{
  \SetScale{#1}
  \begin{axopicture}(70,92)(0,0)
    \Photon[width=.375](35,69)(65,87){3.75}{5}
    \Line[width=.375,dash,dsize=5,arrow](35,69)(5,87)
    \Line[width=.375](35,46)(35,69)
    \Photon[width=.375](35,23)(35,46){3.75}{3}
    \Line[width=1.25](29,46)(41,46)
    \Photon[width=.375](35,23)(65,5){3.75}{5}
    \Photon[width=.375](35,23)(5,5){3.75}{5}
  \end{axopicture}
 }
}
\newcommand{\FGfourGluAuxCanH}[1]{
\hspace{-4pt}
 \raisebox{-.5\height}{
  \SetScale{#1}
  \begin{axopicture}(92,70)(0,0)
    \Photon[width=.34](23,35)(5,5){3}{5}
    \Line[width=.34,dash,dsize=4,arrow](23,35)(5,65)
    \Line[width=.34](23,35)(46,35)    
    \Photon[width=.34](46,35)(69,35){3}{3}
    \Line[width=1](46,30)(46,40)
    \Photon[width=.34](69,35)(87,65){3}{5}
    \Photon[width=.34](69,35)(87,5){3}{5}
  \end{axopicture}
 }
}
\newcommand{\FGfourGluAuxCanX}[1]{
\hspace{-4pt}
 \raisebox{-.5\height}{
  \SetScale{#1}
  \begin{axopicture}(92,70)(0,0)
    \Photon[width=.34](23,35)(87,5){3}{8.3}
    \Line[width=.34,dash,dsize=4,arrow](23,35)(5,65)    
    \Line[width=.34](23,35)(46,35) 
    \Photon[width=.34](46,35)(69,35){3}{3}
    \Line[width=1](46,30)(46,40)
    \Photon[width=.34](69,35)(87,65){3}{5}
    \Photon[width=.34](69,35)(5,5){3}{8.3}
  \end{axopicture}
 }
}

\newcommand{\FGfiveGluAuxCanA}[1]{
\hspace{-4pt}
 \raisebox{-.5\height}{
  \SetScale{#1}
  \begin{axopicture}(100,100)(0,0)
    \Photon[width=.34](83,61)(61,53.5){3}{3.3}
    \Line[width=.34](61,53.5)(50,50)
    \Photon[width=.34](39,46.5)(50,50){3}{1.7}
    \Line[width=1](47.5,57.5)(52.5,42.5)
    \Photon[width=.34](39,46.5)(70.5,22){3}{5}
    \Photon[width=.34](39,46.5)(29.5,22){3}{4}
    \Photon[width=.34](39,46.5)(17,61){3}{4}
    \Bezier[width=.34,dash,dsize=4,arrow](61,53.5)(61,53.5)(50,85)(50,85)
  \end{axopicture}
 }
}
\newcommand{\FGfiveGluAuxCanB}[1]{
\hspace{-4pt}
 \raisebox{-.5\height}{
  \SetScale{#1}
  \begin{axopicture}(100,100)(0,0)
    \Photon[width=.34](70.5,22)(57,41){3}{3.3}
    \Photon[width=.34](43,59)(50,50){3}{1.7}
    \Line[width=.34](50,50)(57,41)
    \Line[width=1](44.375,45.625)(55.625,54.375)
    \Photon[width=.34](43,59)(83,61){3}{6}
    \Photon[width=.34](43,59)(29.5,22){3}{6}
    \Photon[width=.34](43,59)(17,61){3}{4}
    \Bezier[width=.34,dash,dsize=4,arrow,arrowpos=.7](57,41)(61,53.5)(50,85)(50,85)
  \end{axopicture}
 }
}
\newcommand{\FGfiveGluAuxCanC}[1]{
\hspace{-4pt}
 \raisebox{-.5\height}{
  \SetScale{#1}
  \begin{axopicture}(100,100)(0,0)
    \Photon[width=.34](57,59)(83,61){3}{4}
    \Photon[width=.34](57,59)(70.5,22){3}{6}
    \Photon[width=.34](29.5,22)(43,41){3}{3.3}
    \Line[width=.34](43,41)(50,50)
    \Photon[width=.34](50,50)(57,59){3}{1.7}
    \Line[width=1](44.375,54.375)(55.625,45.625)
    \Photon[width=.34](57,59)(17,61){3}{6}
    \Bezier[width=.34,dash,dsize=4,arrow,arrowpos=.7](43,41)(39,54)(39,54)(50,85)
  \end{axopicture}
 }
}
\newcommand{\FGfiveGluAuxCanD}[1]{
\hspace{-4pt}
 \raisebox{-.5\height}{
  \SetScale{#1}
  \begin{axopicture}(100,100)(0,0)
    \Photon[width=.34](17,61)(39,54){3}{3.3}
    \Line[width=.34](50,50)(39,54)
    \Photon[width=.34](61,46)(50,50){3}{1.7}
    \Line[width=1](47.3,42.7)(52.7,57.3)
    \Photon[width=.34](61,46)(83,61){3}{4}
    \Photon[width=.34](61,46)(70.5,22){3}{4}
    \Photon[width=.34](61,46)(29.5,22){3}{6}
    \Bezier[width=.34,dash,dsize=4,arrow](39,54)(39,54)(50,85)(50,85)
  \end{axopicture}
 }
}
\newcommand{\FGghostCanA}[1]{
\hspace{-4pt}
 \raisebox{-.5\height}{
  \SetScale{#1}
  \begin{axopicture}(92,70)(0,0)
    \Line[width=.34,dash,dsize=4,arrow](23,35)(5,5)
    \Line[arrow,arrowpos=0.7,arrowlength=7.5,arrowwidth=5,arrowinset=0](0,73.5)(5,65)
    \Photon[width=.34](23,35)(5,65){3}{5}
    \Line[width=.34,dash,dsize=4,arrow](69,35)(23,35)
    \Line[width=.34,dash,dsize=4,arrow](87,65)(69,35)
    \Photon[width=.34](69,35)(87,5){3}{5}
  \end{axopicture}
 }
}
\newcommand{\FGghostCanB}[1]{
\hspace{-4pt}
 \raisebox{-.5\height}{
  \SetScale{#1}
  \begin{axopicture}(92,70)(0,0)
    \Line[arrow,arrowpos=0.7,arrowlength=7.5,arrowwidth=5,arrowinset=0](0,-3.5)(5,5)
    \Photon[width=.34](23,35)(5,5){3}{5}
    \Line[width=.34,dash,dsize=4,arrow](23,35)(5,65)
    \Line[width=.34,dash,dsize=4,arrow](69,35)(23,35)
    \Line[width=.34,dash,dsize=4,arrow](87,65)(69,35)
    \Photon[width=.34](69,35)(87,5){3}{5}
  \end{axopicture}
 }
}
\newcommand{\FGghostCanC}[1]{
\hspace{-4pt}
 \raisebox{-.5\height}{
  \SetScale{#1}
  \begin{axopicture}(70,92)(0,0)
    \Bezier[width=.375,dash,dsize=5,arrow,arrowpos=0.7](35,69)(20,64)(5,5)(5,5)
    \Line[width=.375,dash,dsize=5,arrow](65,87)(35,69)
    \Photon[width=.375](35,46)(35,69){3.75}{3}
    \Line[width=.375](35,23)(35,46)
    \Line[width=1.25](29,46)(41,46)
    \Bezier[width=.375,dash,dsize=5,arrow,arrowpos=0.7](35,23)(20,28)(5,87)(5,87)
    \Photon[width=.375](35,23)(65,5){3.75}{5}
  \end{axopicture}
 }
}
\newcommand{\FGghostCanD}[1]{
\hspace{-4pt}
 \raisebox{-.5\height}{
  \SetScale{#1}
  \begin{axopicture}(70,92)(0,0)
    \Line[width=.375,dash,dsize=5,arrow](35,69)(5,87)
    \Line[width=.375,dash,dsize=5,arrow](65,87)(35,69)
    \Photon[width=.375](35,46)(35,69){3.75}{3}
    \Line[width=.375](35,23)(35,46)
    \Line[width=1.25](29,46)(41,46)
    \Line[width=.375,dash,dsize=5,arrow](35,23)(5,5)
    \Photon[width=.375](35,23)(65,5){3.75}{5}
  \end{axopicture}
 }
}
\newcommand{\FGggPropContrCan}[1]{
\hspace{-4pt}
 \raisebox{-.5\height}{
  \SetScale{#1}
  \begin{axopicture}(90,20)(0,40)
    \DoublePhoton(20,50)(80,50){4}{5}{1.5}
    \Line[arrow,arrowpos=0.6,arrowlength=11.25,arrowwidth=7.5,arrowinset=0](10,50)(20,50)
  \end{axopicture}
 }
}
\newcommand{\FGhhPropContrCan}[1]{
\hspace{-4pt}
 \raisebox{-.5\height}{
  \SetScale{#1}
  \begin{axopicture}(90,20)(0,40)
    \Line[arrow,dash,dsize=6,arrowscale=1.5](70,50)(10,50)
    \DoublePhoton(80,50)(90,50){4}{1}{1.5}
\Line[arrow,arrowpos=0.5,arrowlength=11.25,arrowwidth=7.5,arrowinset=0,arrowstroke=.5](70,50)(80,50)
  \end{axopicture}
 }
}
\newcommand{\FGgggContrCan}[1]{
\hspace{-4pt}
 \raisebox{-.5\height}{
  \SetScale{#1}
  \begin{axopicture}(70,70)(15,15)
    \DoublePhoton[width=.34](50,50)(50,85){3}{5}{1}
    \Line[arrow,arrowpos=0.7,arrowlength=7.5,arrowwidth=5,arrowinset=0](50,95)(50,85)
    \DoublePhoton[width=.34](50,50)(80,32){3}{5}{1}
    \DoublePhoton[width=.34](50,50)(20,32){3}{5}{1}
    \end{axopicture}
 }
}
\newcommand{\FGauxVertexCanL}[1]{
\hspace{-4pt}
 \raisebox{-.5\height}{
  \SetScale{#1}
  \begin{axopicture}(70,70)(15,15)
    \Line[width=.34,dash,dsize=4,arrow](50,50)(50,85)
    \DoublePhoton[width=.34](50,50)(80,32){3}{5}{1}
    \DoublePhoton[width=.34](20,32)(35,41){3}{2.5}{1}
    \Line[width=.34](50,50)(35,41)
    \Line[width=1](32,46)(38,36)
  \end{axopicture}
 }
}
\newcommand{\FGghostVertexCanL}[1]{
\hspace{-4pt}
 \raisebox{-.5\height}{
  \SetScale{#1}
  \begin{axopicture}(70,70)(15,15)
    \Line[width=.34,dash,dsize=4,arrow](50,50)(50,85)
    \Line[width=.34,dash,dsize=4,arrow](20,32)(50,50)
    \DoublePhoton[width=.34](50,50)(80,32){3}{5}{1}
    \DoublePhoton[width=.34](14,28)(20,32){3}{1}{1}
\Line[width=.34,arrow,arrowpos=0.1,arrowlength=7.5,arrowwidth=5,arrowinset=0,arrowstroke=.34](23,
34)(20,32)
  \end{axopicture}
 }
}
\newcommand{\FGauxVertexCanR}[1]{
\hspace{-4pt}
 \raisebox{-.5\height}{
  \SetScale{#1}
  \begin{axopicture}(70,70)(15,15)
    \Line[width=.34,dash,dsize=4,arrow](50,50)(50,85)
    \DoublePhoton[width=.34](80,32)(65,41){3}{2.5}{1}
    \Line[width=.34](50,50)(65,41)
    \Line[width=1](62,36)(68,46)
    \DoublePhoton[width=.34](50,50)(20,32){3}{5}{1}
  \end{axopicture}
 }
}
\newcommand{\FGghostVertexCanR}[1]{
\hspace{-4pt}
 \raisebox{-.5\height}{
  \SetScale{#1}
  \begin{axopicture}(70,85)(15,15)
    \Line[width=.34,dash,dsize=4,arrow](50,50)(50,85)
    \Line[width=.34,dash,dsize=4,arrow](80,32)(50,50)
    \DoublePhoton[width=.34](86,28)(80,32){3}{1}{1}
\Line[width=.34,arrow,arrowpos=0.1,arrowlength=7.5,arrowwidth=5,arrowinset=0,arrowstroke=.34](77,
34)(80,32)
    \DoublePhoton[width=.34](50,50)(20,32){3}{5}{1}
  \end{axopicture}
 }
}
\newcommand{\FGfourGContrCan}[1]{
\hspace{-4pt}
 \raisebox{-.5\height}{
  \SetScale{#1}
  \begin{axopicture}(70,100)(15,0)
    \DoublePhoton[width=.34](50,50)(75,75){3}{5}{1}
    \DoublePhoton[width=.34](50,50)(75,25){3}{5}{1}
    \DoublePhoton[width=.34](50,50)(25,25){3}{5}{1}
    \DoublePhoton[width=.34](50,50)(25,75){3}{5}{1}
    \Line[arrow,arrowpos=0.7,arrowlength=7.5,arrowwidth=5,arrowinset=0](18,82)(25,75)
  \end{axopicture}
 }
}
\newcommand{\FGfourGauxCanI}[1]{
\hspace{-4pt}
 \raisebox{-.5\height}{
  \SetScale{#1}
  \begin{axopicture}(70,92)(0,0)
    \DoublePhoton[width=.375](35,69)(65,87){3.75}{5}{1.125}
    \Line[width=.375,dash,dsize=5,arrow](35,69)(5,87)
    \Line[width=.375](35,46)(35,69)
    \DoublePhoton[width=.375](35,23)(35,46){3.75}{3}{1.125}
    \Line[width=1.25](29,46)(41,46)
    \DoublePhoton[width=.375](35,23)(65,5){3.75}{5}{1.125}
    \DoublePhoton[width=.375](35,23)(5,5){3.75}{5}{1.125}
  \end{axopicture}
 }
}
\newcommand{\FGfourGauxCanH}[1]{
\hspace{-4pt}
 \raisebox{-.5\height}{
  \SetScale{#1}
  \begin{axopicture}(92,70)(0,0)
    \DoublePhoton[width=.34](23,35)(5,5){3}{5}{1}
    \Line[width=.34,dash,dsize=4,arrow](23,35)(5,65)
    \Line[width=.34](23,35)(46,35)    
    \DoublePhoton[width=.34](46,35)(69,35){3}{3}{1}
    \Line[width=1](46,30)(46,40)
    \DoublePhoton[width=.34](69,35)(87,65){3}{5}{1}
    \DoublePhoton[width=.34](69,35)(87,5){3}{5}{1}
  \end{axopicture}
 }
}
\newcommand{\FGfourGauxCanX}[1]{
\hspace{-4pt}
 \raisebox{-.5\height}{
  \SetScale{#1}
  \begin{axopicture}(92,70)(0,0)
    \DoublePhoton[width=.34](23,35)(87,5){3}{8.3}{1}
    \Line[width=.34,dash,dsize=4,arrow](23,35)(5,65)    
    \Line[width=.34](23,35)(46,35) 
    \DoublePhoton[width=.34](46,35)(69,35){3}{3}{1}
    \Line[width=1](46,30)(46,40)
    \DoublePhoton[width=.34](69,35)(87,65){3}{5}{1}
    \DoublePhoton[width=.34](69,35)(5,5){3}{8.3}{1}
  \end{axopicture}
 }
}
\newcommand{\FGfiveGContrCan}[1]{
\hspace{-4pt}
 \raisebox{-.5\height}{
  \SetScale{#1}
  \begin{axopicture}(100,100)(0,0)
    \DoublePhoton[width=.34](50,50)(50,85){3}{5}{1}
    \DoublePhoton[width=.34](50,50)(83,61){3}{5}{1}
    \DoublePhoton[width=.34](50,50)(70.5,22){3}{5}{1}
    \DoublePhoton[width=.34](50,50)(29.5,22){3}{5}{1}
    \DoublePhoton[width=.34](50,50)(17,61){3}{5}{1}
    \Line[arrow,arrowpos=0.7,arrowlength=7.5,arrowwidth=5,arrowinset=0](50,95)(50,85)
  \end{axopicture}
 }
}
\newcommand{\FGfiveGauxCanA}[1]{
\hspace{-4pt}
 \raisebox{-.5\height}{
  \SetScale{#1}
  \begin{axopicture}(100,100)(0,0)
    \DoublePhoton[width=.34](83,61)(61,53.5){3}{3.3}{1}
    \Line[width=.34](61,53.5)(50,50)
    \DoublePhoton[width=.34](39,46.5)(50,50){3}{1.7}{1}
    \Line[width=1](47.5,57.5)(52.5,42.5)
    \DoublePhoton[width=.34](39,46.5)(70.5,22){3}{5}{1}
    \DoublePhoton[width=.34](39,46.5)(29.5,22){3}{4}{1}
    \DoublePhoton[width=.34](39,46.5)(17,61){3}{4}{1}
    \Bezier[width=.34,dash,dsize=4,arrow](61,53.5)(61,53.5)(50,85)(50,85)
  \end{axopicture}
 }
}
\newcommand{\FGfiveGauxCanB}[1]{
\hspace{-4pt}
 \raisebox{-.5\height}{
  \SetScale{#1}
  \begin{axopicture}(100,100)(0,0)
    \DoublePhoton[width=.34](70.5,22)(57,41){3}{3.3}{1}
    \DoublePhoton[width=.34](43,59)(50,50){3}{1.7}{1}
    \Line[width=.34](50,50)(57,41)
    \Line[width=1](44.375,45.625)(55.625,54.375)
    \DoublePhoton[width=.34](43,59)(83,61){3}{6}{1}
    \DoublePhoton[width=.34](43,59)(29.5,22){3}{6}{1}
    \DoublePhoton[width=.34](43,59)(17,61){3}{4}{1}
    \Bezier[width=.34,dash,dsize=4,arrow,arrowpos=.7](57,41)(61,53.5)(50,85)(50,85)
  \end{axopicture}
 }
}
\newcommand{\FGfiveGauxCanC}[1]{
\hspace{-4pt}
 \raisebox{-.5\height}{
  \SetScale{#1}
  \begin{axopicture}(100,100)(0,0)
    \DoublePhoton[width=.34](57,59)(83,61){3}{4}{1}
    \DoublePhoton[width=.34](57,59)(70.5,22){3}{6}{1}
    \DoublePhoton[width=.34](29.5,22)(43,41){3}{3.3}{1}
    \Line[width=.34](43,41)(50,50)
    \DoublePhoton[width=.34](50,50)(57,59){3}{1.7}{1}
    \Line[width=1](44.375,54.375)(55.625,45.625)
    \DoublePhoton[width=.34](57,59)(17,61){3}{6}{1}
    \Bezier[width=.34,dash,dsize=4,arrow,arrowpos=.7](43,41)(39,54)(39,54)(50,85)
  \end{axopicture}
 }
}
\newcommand{\FGfiveGauxCanD}[1]{
\hspace{-4pt}
 \raisebox{-.5\height}{
  \SetScale{#1}
  \begin{axopicture}(100,100)(0,0)
    \DoublePhoton[width=.34](17,61)(39,54){3}{3.3}{1}
    \Line[width=.34](50,50)(39,54)
    \DoublePhoton[width=.34](61,46)(50,50){3}{1.7}{1}
    \Line[width=1](47.3,42.7)(52.7,57.3)
    \DoublePhoton[width=.34](61,46)(83,61){3}{4}{1}
    \DoublePhoton[width=.34](61,46)(70.5,22){3}{4}{1}
    \DoublePhoton[width=.34](61,46)(29.5,22){3}{6}{1}
    \Bezier[width=.34,dash,dsize=4,arrow](39,54)(39,54)(50,85)(50,85)
  \end{axopicture}
 }
}
\newcommand{\FGsixGContrCan}[1]{
\hspace{-4pt}
 \raisebox{-.5\height}{
  \SetScale{#1}
  \begin{axopicture}(80,100)(10,0)
    \DoublePhoton[width=.34](50,50)(50,85){3}{5}{1}
    \DoublePhoton[width=.34](50,50)(82,67.5){3}{5}{1}
    \DoublePhoton[width=.34](50,50)(82,32.5){3}{5}{1}
    \DoublePhoton[width=.34](50,50)(50,15){3}{5}{1}
    \DoublePhoton[width=.34](50,50)(18,32.5){3}{5}{1}
    \DoublePhoton[width=.34](50,50)(18,67.5){3}{5}{1}
    \Line[arrow,arrowpos=0.7,arrowlength=7.5,arrowwidth=5,arrowinset=0](50,95)(50,85)
  \end{axopicture}
 }
}
\newcommand{\FGsixGauxCanA}[1]{
\hspace{-4pt}
 \raisebox{-.5\height}{
  \SetScale{#1}
  \begin{axopicture}(80,100)(10,0)
    \DoublePhoton[width=.34](82,67.5)(61,56){3}{3.3}{1}
    \DoublePhoton[width=.34](40,44)(50.5,50){3}{1.7}{1}
    \Line[width=.34](50.5,50)(61,56)
    \Line[width=1](47,56.5)(54,43.5)
    \DoublePhoton[width=.34](40,44)(82,32.5){3}{6}{1}
    \DoublePhoton[width=.34](40,44)(50,15){3}{4.5}{1}
    \DoublePhoton[width=.34](40,44)(18,32.5){3}{3.3}{1}
    \DoublePhoton[width=.34](40,44)(18,67.5){3}{5}{1}
    \Bezier[width=.34,dash,dsize=4,arrow](61,56)(61,56)(50,85)(50,85)
  \end{axopicture}
 }
}
\newcommand{\FGsixGauxCanB}[1]{
\hspace{-4pt}
 \raisebox{-.5\height}{
  \SetScale{#1}
  \begin{axopicture}(80,100)(10,0)
    \DoublePhoton[width=.34](39,56)(82,67.5){3}{6}{1}
    \DoublePhoton[width=.34](39,56)(50,50){3}{1.7}{1}
    \Line[width=.34](50,50)(61,44)
    \Line[width=1](53.5,56.5)(46.5,43.5)
    \DoublePhoton[width=.34](39,56)(50,15){3}{6}{1}
    \DoublePhoton[width=.34](39,56)(18,32.5){3}{5}{1}
    \DoublePhoton[width=.34](39,56)(18,67.5){3}{3.3}{1}
    \DoublePhoton[width=.34](82,32.5)(61,44){3}{3.3}{1}
    \Bezier[width=.34,dash,dsize=4,arrow,arrowpos=.7](61,44)(61,44)(61,56)(50,85)
  \end{axopicture}
 }
}
\newcommand{\FGsixGauxCanC}[1]{
\hspace{-4pt}
 \raisebox{-.5\height}{
  \SetScale{#1}
  \begin{axopicture}(80,100)(10,0)
    \DoublePhoton[width=.34](50,55)(82,67.5){3}{4.7}{1}
    \DoublePhoton[width=.34](50,55)(82,32.5){3}{6}{1}
    \DoublePhoton[width=.34](50,35)(50,15){3}{3.3}{1}
    \DoublePhoton[width=.34](50,45)(50,55){3}{1.7}{1}
    \Line[width=.34](50,45)(50,35)
    \Line[width=1](45,45)(55,45)
    \DoublePhoton[width=.34](50,55)(18,32.5){3}{6}{1}
    \DoublePhoton[width=.34](50,55)(18,67.5){3}{4.7}{1}
    \Bezier[width=.34,dash,dsize=4,arrow,arrowpos=.7](50,35)(61,44)(61,56)(50,85)
  \end{axopicture}
 }
}
\newcommand{\FGsixGauxCanD}[1]{
\hspace{-4pt}
 \raisebox{-.5\height}{
  \SetScale{#1}
  \begin{axopicture}(80,100)(10,0)
    \DoublePhoton[width=.34](61,56)(82,67.5){3}{3.3}{1}
    \DoublePhoton[width=.34](61,56)(50.5,50){3}{1.7}{1}
    \Line[width=.34](40,44)(50.5,50)
    \Line[width=1](47,56.5)(54,43.5)
    \DoublePhoton[width=.34](61,56)(82,32.5){3}{4.5}{1}
    \DoublePhoton[width=.34](61,56)(50,15){3}{6}{1}
    \DoublePhoton[width=.34](18,32.5)(40,44){3}{3.3}{1}
    \DoublePhoton[width=.34](61,56)(18,67.5){3}{6}{1}
    \Bezier[width=.34,dash,dsize=4,arrow,arrowpos=.7](40,44)(40,44)(39,56)(50,85)
  \end{axopicture}
 }
}
\newcommand{\FGsixGauxCanE}[1]{
\hspace{-4pt}
 \raisebox{-.5\height}{
  \SetScale{#1}
  \begin{axopicture}(80,100)(10,0)
    \DoublePhoton[width=.34](61,44)(82,67.5){3}{5}{1}
    \DoublePhoton[width=.34](61,44)(82,32.5){3}{3.3}{1}
    \DoublePhoton[width=.34](61,44)(50,15){3}{5}{1}
    \DoublePhoton[width=.34](61,44)(18,32.5){3}{6}{1}
    \DoublePhoton[width=.34](61,44)(50,50){3}{1.7}{1}
    \Line[width=.34](50,50)(39,56)
    \Line[width=1](53.5,56.5)(46.5,43.5)
    \DoublePhoton[width=.34](18,67.5)(39,56){3}{3.3}{1}
    \Line[width=.34,dash,dsize=4,arrow](39,56)(50,85)
  \end{axopicture}
 }
}
\newcommand{\FGghostGCanA}[1]{
\hspace{-4pt}
 \raisebox{-.5\height}{
  \SetScale{#1}
  \begin{axopicture}(92,70)(0,0)
    \Line[width=.34,dash,dsize=4,arrow](23,35)(5,5)
    \Line[arrow,arrowpos=0.7,arrowlength=7.5,arrowwidth=5,arrowinset=0,arrowstroke=.5](0,73.5)(5,65)
    \DoublePhoton[width=.34](23,35)(5,65){3}{5}{1}
    \Line[width=.34,dash,dsize=4,arrow](69,35)(23,35)
    \Line[width=.34,dash,dsize=4,arrow](87,65)(69,35)
    \DoublePhoton[width=.34](69,35)(87,5){3}{5}{1}
  \end{axopicture}
 }
}
\newcommand{\FGghostGCanB}[1]{
\hspace{-4pt}
 \raisebox{-.5\height}{
  \SetScale{#1}
  \begin{axopicture}(92,70)(0,0)
    \Line[arrow,arrowpos=0.7,arrowlength=7.5,arrowwidth=5,arrowinset=0,arrowstroke=.5](0,-3.5)(5,5)
    \DoublePhoton[width=.34](23,35)(5,5){3}{5}{1}
    \Line[width=.34,dash,dsize=4,arrow](23,35)(5,65)
    \Line[width=.34,dash,dsize=4,arrow](69,35)(23,35)
    \Line[width=.34,dash,dsize=4,arrow](87,65)(69,35)
    \DoublePhoton[width=.34](69,35)(87,5){3}{5}{1}
  \end{axopicture}
 }
}
\newcommand{\FGghostGCanC}[1]{
\hspace{-4pt}
 \raisebox{-.5\height}{
  \SetScale{#1}
  \begin{axopicture}(70,92)(0,0)
    \Bezier[width=.375,dash,dsize=5,arrow,arrowpos=0.7](35,69)(20,64)(5,5)(5,5)
    \Line[width=.375,dash,dsize=5,arrow](65,87)(35,69)
    \DoublePhoton[width=.375](35,46)(35,69){3.75}{3}{1.125}   
    \Line[width=.375](35,23)(35,46)
    \Line[width=1.25](29,46)(41,46)
    \Bezier[width=.375,dash,dsize=5,arrow,arrowpos=0.7](35,23)(20,28)(5,87)(5,87)
    \DoublePhoton[width=.375](35,23)(65,5){3.75}{5}{1.125}
  \end{axopicture}
 }
}
\newcommand{\FGghostGCanD}[1]{
\hspace{-4pt}
 \raisebox{-.5\height}{
  \SetScale{#1}
  \begin{axopicture}(70,92)(0,0)
    \Line[width=.375,dash,dsize=5,arrow](35,69)(5,87)
    \Line[width=.375,dash,dsize=5,arrow](65,87)(35,69)
    \DoublePhoton[width=.375](35,46)(35,69){3.75}{3}{1.125}   
    \Line[width=.375](35,23)(35,46)
    \Line[width=1.25](29,46)(41,46)
    \Line[width=.375,dash,dsize=5,arrow](35,23)(5,5)
    \DoublePhoton[width=.375](35,23)(65,5){3.75}{5}{1.125}
  \end{axopicture}
 }
}

\begin{document}

\begin{frontmatter}
 
\title{Off-shell Diagrammatics for Quantum Gravity}

\author{Henry Kißler}
\ead{kissler@physik.hu-berlin.de}
\address{Department of Mathematics, Humboldt-Universität zu Berlin, Rudower Chaussee 25, 
D-12489 Berlin, Germany}
\begin{abstract}
This article reports on how diagrammatic identities of Yang--Mills theory translate to 
diagrammatics for pure gravity. For this, we consider the Einstein--Hilbert action and follow the 
approach of Capper, Leibbrandt, and Medrano and expand the inverse metric density around the 
Minkowski metric. By analogy to Yang--Mills theory, cancellation identities are constructed 
for the graviton as well as the ghost vertices up to the valency of six.
\end{abstract}

\end{frontmatter}

\section{Introduction}

Ideas and methods from algebraic geometry have been shown to successfully inform our understanding 
of perturbative quantum field theory 
\cite{Bloch:2005bh,Panzer:2015ida,Bloch:2015efx,Brown:2015fyf,Schnetz:2017bko,Schnetz:2019cab,
Panzer:2019yxl,Klausen:2019hrg,Tapuskovic:2019cpr}. Most of these mathematical methods utilize the 
parametric representation of Feynman rules that is based on two graph polynomials, usually called 
Kirchhoff or Symanzik polynomials. In \cite{Kreimer:2012jw} a third graph polynomial has been 
introduced in order to obtain a manifest representation of tensorial amplitudes in Yang--Mills 
theory. This polynomial is called corolla polynomial for its parameters are indexed by 
the half-edges of a 3-valent graph $G$. By acting with the corolla polynomial $C(G)$ as a 
differential operator on the scalar integrand of $G$, all contributions of Yang--Mills diagrams are 
obtained that can be constructed from $G$ by collapsing some edges that connect two 3-valent 
vertices or by turning cycles into ghost cycles (in this case both orientations for each ghost 
cycle are considered). These two diagrammatic operations, edge-collapsing and generating ghost 
cycles, turn out to be useful, because they constitute two cochain complexes, which are called the 
gauge and ghost cycle complexes \cite{Kreimer:2012jw,Berghoff2020}. Crucially, the physical 
amplitudes lie in the kernel of these operations, which is similar to the BRST complex. The exact 
relation between BRST and these gauge complexes as well as concrete physical constrains are an 
interesting topic of ongoing research. 

Aside from that, it has been shown that the representation in terms of graph polynomials can be 
extended to the full Standard Model of elementary particle physics \cite{Prinz:2016fka}. The 
natural question whether this approach extends to a quantum theory of gravity remains to be 
answered. This question, however, is much more intricate due to the infinite number of 
gravitational vertices and their algebraic complexity. As a first step into this direction, the 
momentum space Feynman rules of gravity have been derived combinatorially for any valence in 
\cite{Cheung:2017kzx,Prinz:2018dhq,Prinz:2020nru}. We are approaching this topic by a different 
angle, namely we 
are pointing out how the diagrammatic cancellation identities of Yang--Mills theory translate to 
quantum gravity when quantized following Capper, Leibbrandt, and Medrano \cite{Capper:1973pv}. The 
high degree of similarity between Yang--Mills and gravity ought to allow for a deduction of 
appropriate cochain maps and their complexes for gravity. We believe that both such a construction 
and the constraints, that result from identities reported here, inform the construction of a 
corolla polynomial for gravity.

\section{Yang--Mills theory}

This section reviews the cancellation identities underlying Yang--Mills theory, because the 
subsequent construction of gravitational identities will depend on them.

\subsection{Model and conventions}
Consider the standard Yang--Mills Lagrangian with a linear covariant gauge fixing
\begin{align}
 \lagrangian = & \lagrangian_\textnormal{Dirac} + \lagrangian_\textnormal{YM} + 
\lagrangian_\textnormal{gf} + \frac{1}{ξ}\lagrangian_\text{gh} \label{eq:qcdLagrangian}\\
 & \quad \lagrangian_\textnormal{Dirac} = i \psibar \slashed{D} ψ,\quad \textnormal{where} \ 
 \left(D_μ\right)_{jk} = δ_{jk} \del_μ - i g T^a_{jk} A_μ^a\\
 & \quad \lagrangian_\textnormal{YM} = - \frac{1}{4} F_{μν}^{a} F^{μν a},\quad \textnormal{where} \ 
 F_{μν}^a = \del_μ A_ν^a - \del_ν A_μ^a + g \f{a b c} A^b_μ A^c_ν\\
 & \quad \lagrangian_\textnormal{gf} = - \frac{1}{2ξ} (\partial^μA^a_{μ}) (\partial^νA^a_ν), \quad 
\phantom{\textnormal{where}} \ \lagrangian_\textnormal{gh} = - 
(\partial^μ\cbar^a)\left[δ_{ac}\partial_μ + g f^{abc}A^b_μ\right]c^c.
\end{align}
Here we mainly follow the conventions of \cite{Pascual:1984zb} with the exception of an additional 
factor of $\nicefrac{1}{ξ}$ in front of the ghost Lagrangian $\lagrangian_{\textnormal{gh}}$. By 
means of standard quantization procedures and perturbative expansion, the well-known Feynman 
rules of this model are readily derived. For the subsequent discussion, we are mainly concerned 
with the diagrammatic expressions. The residues (i.e.\ edges and vertices; see 
\cite{Kreimer:2012jw,Kreimer:2005rw}) of this model read
\begin{align}
 \mathcal{R}_\textnormal{YM} = \Bigg\{\FGqqProp{.75}, \FGgluProp{.75}, 
\FGhhProp{.75},\FGqqGluVertex{.75}, \FGtriGluVertex{.75}, \FGtetraGluVertex{.75}, 
\FGhhGluVertex{.75}\Bigg\}.
\end{align}
As usual, quarks are represented by straight directed lines, the wavy lines represent gluons and 
the dashed directed lines represent the ghosts. For reasons that will be clear later on, we will 
restrict our attention to residues without quarks in the following.

\subsection{Slavnov--Taylor identities}

The enormous phenomenological success of Yang--Mills theories in the high-energy domain is 
crucially founded on what is called renormalizability. With an appropriate treatment of 
singularities, which are based in quantum corrections, renormalizability translates into the famous 
Slavnov--Taylor identities \cite{Slavnov:1972fg,Taylor:1971ff}. For instance, the respective 
identity for the 1-particle irreducible 3-point function at first loop order is
\begin{align}
 \FGtriGluContrBlob{.75} = \FGtriGluTransLBlob{.75} + \FGtriGluTransRBlob{.75} 
 - \FGtriGluPropLBlob{.75} - \FGtriGluPropRBlob{.75} - \FGtriGluGhBlob{.75}, 
\end{align}
where the blob represents all 1-particle irreducible diagrams. Furthermore, the slashed edges, 
which pair the straight with the wavy half-edges, have been introduced to represent 
cancelled gluon propagators and the triangle on the left-hand side and letters $\textnormal{T}$ 
represent certain projectors acting on the external gluons. For further details, the interested 
reader is referred to \cite{Gracey:2019mix} where this identity has not only been derived by both 
Hopf-algebraic and diagrammatic methods, but has also been perturbatively verified in dimensional 
regularization in the most general momentum setting. Here, it is worth emphasizing that the 
diagrammatic technique solely relies on a set of cancellation identities at tree--level, which will 
be discussed in the following section.

\subsection{Cancellation identities}

Most of the cancellation identities shown in this sections go back to the early work of Lautrup and 
Cvitanović \cite{Lautrup:1977ty,Cvitanovic:1977qu}. Since these will establish the foundation for 
our subsequent construction of gravity cancellations, we need to review them in some detail. Here, 
we will mainly focus on the diagrammatics. A reader interested in the analytic details might want to
consider \cite{Gracey:2019mix} to complement the original work of Lautrup and Cvitanović. The first 
identity relates the gluon propagator to the ghost propagator
\begin{align}
\label{eq:canPropYM}
  \FGgluPropContrCan{.5} & = \FGghPropContrCan{.5}.\\
\intertext{Here, the black triangle represents a longitudinal contraction of the gluon propagator. 
To be more precise, abbreviating the gluon propagator by $D^{μν}_{ab}(p)$ (again we follow the 
conventions of \cite{Pascual:1984zb}, but drop the additional superscript $0$), the left hand side 
represents the longitudinal contraction, that is $p_μ 
D^{μν}_{ab}(p)$. An easy derivation shows that contracted propagator equals the ghost propagator 
times the longitudinal contraction with respect to the remaining Lorentz index $ν$, explaining 
the black triangle on the right. Similar identities hold for the vertices. A longitudinal 
contraction of a 3-valent gluon vertex yields}
\label{eq:canTriYM}
  \FGtriGluContrCan{.75} & = 
  \FGauxiVertexCanL{.75}+\FGghVertexCanL{.75}+\FGauxiVertexCanR{.75}+\FGghVertexCanR{.75}.\\
\intertext{Here, we needed to introduce an auxiliary vertex that is connected by a wavy gluon line, 
a straight line and an outgoing ghost line. Labelling the straight and wavy lines with Lorentz 
indices $μ$ and $ν$ respectively, the Feynman rule of this auxiliary vertex equals (up to a 
conventional complex constant) to Minkowski metric $η^{μν}$ and a factor of the structure 
constants. Also note the slashed propagators in the first and third diagrams that connect a straight 
to the wavy gluon line. These slashed propagators represent the inverse of the gluon propagator and 
result in contractions as utilized in \cite{Kreimer:2012jw}. The other diagrams incorporate the 
usual ghost--gluon vertex followed by an ongoing longitudinal contraction. Similarly, one shows the 
following identities for the 4-valent vertex.}
\label{eq:canTetraYM}
  \FGfourGluContrCan{.75} &= - \FGfourGluAuxCanI{.66} - \FGfourGluAuxCanH{.75} - 
  \FGfourGluAuxCanX{.75}.\\
\label{eq:canPentaYM}
 0 \quad &= \FGfiveGluAuxCanA{.75} + \FGfiveGluAuxCanB{.75} +\FGfiveGluAuxCanC{.75} 
+\FGfiveGluAuxCanD{.75}\\
\intertext{Their similarity to Lie algebra cohomology, namely the IHX relation, is not by accident, 
but due to a Jacobi identity for the structure constants for the underlying Lie algebra of the 
gauge group. Finally, the cancellation of ghost loops is explained by the identity}
\label{eq:canGhostYM}
 0 \quad &= \FGghostCanA{.75} - \FGghostCanB{.75} + \FGghostCanC{.75} - \FGghostCanD{.75}.
\end{align}

\section{Pure gravity}

Gravity was shown to be not renormalizable in the traditional sense in the background field gauge 
\cite{tHooft:1974toh,Goroff:1985sz,Goroff:1985th} and heat-kernel method \cite{vandeVen:1991gw}. 
Nonetheless, renormalization is possible when understood as an effective field theory 
\cite{Weinberg:2009bg,Weinberg:2016kyd} by adding higher derivative terms to the Lagrangian. This 
situation raises the question what kind of structure these additional terms obey and how they can be 
characterized. In the following, we first introduce the model in a tensor-density-based approach 
and secondly report on cancellation relations for all vertices that might be encountered in 
propagator corrections at two loops, which is know to be the lowest order to require 
subtraction of higher derivative terms.

\subsection{The model}

The most established way to quantize the Einstein--Hilbert action utilizes the background field 
technique. Crucially, the metric tensor is decomposed into a combination of a classical background 
metric and a quantum field. The caveat of the approach is that the expansion of the Lagrangian 
yields a term which only depends linearly on the quantum metric. In perturbation theory, this term 
corresponds to a non-interacting source of the graviton field that results in superfluous 
complications in Feynman diagrammatic computations. Therefore, these linear terms are usually 
removed by somehow artificial subtraction.

Here, we follow a complementary approach that consists in quantizing the so-called tensor density 
instead of the metric tensor which goes back to \cite{Capper:1973pv} and is also known as the 
Landau-Lifshitz approach in the context of classical field theory \cite{Landau:1951dk}. First, the 
tensor density $\widetilde{g}$ is defined to be the metric tensor $g$ scaled by the inverse of 
local coordinate function of the Riemannian volume form. In the following, the Einstein--Hilbert 
action is understood as a function of the tensor density rather than the metric. Also, we choose 
to quantize the inverse tensor density rather then just the tensor in the customary quantization 
procedure. Therefore, the action is perturbatively expanded in terms of the inverse tensor density
\begin{align}
\label{eq:expandMetric}
 \widetilde{g}^{μν} = \sqrt{-|g|}g^{μν} = η^{μν} + κ Φ^{μν}
\end{align}
which decomposes into the Minkowski metric $η$ and the quantum, also called graviton, field $Φ$. As 
it was observed in \cite{Capper:1973pv}, this procedure avoids the tedious source terms of the 
graviton field; the authors defined a simplified version of the gauge harmonic gauge fixing and 
derived an appropriate ghost Lagrangian. In this work, we will mainly follow their conventions 
throughout:
\begin{align}
 i S = & i \int d^4x \left( \lagrangian_{\textnormal{EH}} + \lagrangian_{\textnormal{gf}} + 
\frac{1}{ξ}\lagrangian_{\textnormal{gh}} \right)\\
\label{eq:EHgaugeFixing}
& \quad \lagrangian_{\textnormal{EH}} = - \frac{2}{κ^2} R \sqrt{-|g|}, \quad\quad
\lagrangian_{\textnormal{gf}} = - \frac{1}{2ξ} η_{νσ} (\partial_μ Φ^{μν}) (\partial_ρ Φ^{ρσ}), 
\quad\quad \lagrangian_{\textnormal{gh}} = \cbar_μ \partial_ν D^{μνλ} c_λ,\\
& \quad D^{μνλ} =  η^{μλ}\partial^ν + κ\left[ - (\partial^λ Φ^{μν}) - Φ^{μν} \partial^λ + 
η^{μλ}Φ^{νρ} \partial_ρ + η^{νλ} Φ^{μρ}\partial_ρ\right].
\end{align}
For a detailed account on the construction of the ghost Lagrangian and derivation of Feynman rules 
the reader is referred to \cite{Prinz:2020nru}. A major advantage of this approach is that the 
perturbative expansion of the Einstein--Hilbert Lagrangian is free of linear graviton terms. In 
addition to that, the Feynman rules of the propagators and the 3-valent vertex are available in 
the literature 
\cite{Capper:1978yf} allowing for cross checks of our results.

In this conventions, the action gives rise to the residues
\begin{align}
\mathcal{R}_\textnormal{EH} = \Bigg\{\FGggProp{.75}, \FGhhProp{.75}, \FGgggVertex{.75}, 
\FGhhgVertex{.75}, \FGggggVertex{.75}, \FGgggggVertex{.75},  \FGggggggVertex{.75}, \dotsc \Bigg\},
\end{align}
where it is worth noting that the expansion of the inverse metric \eqref{eq:expandMetric} in 
combination with the simple gauge fixing ensures that the interaction of graviton and their ghosts 
is due to a single 3-valent vertex. We use {\sc FORM} \cite{Vermaseren:2000nd,Ruijl:2017dtg} to 
expand the action and derive Feynman rules up to the 6-valent graviton vertex. Technically, our 
implementation allows for higher orders as well, but we stop at the 6-valent vertex for practical 
reasons---these vertices are sufficient to study the two-point function at two loops, which is 
expected to be the lowest order that requires higher derivative counterterms and hence reveals the 
non-renormalizability aspect of gravity.

\subsection{Cancellation identities}

Subsequently, we take Kreimer's analogy \cite{Kreimer:2008dg} of gravity and non-abelian gauge 
theories seriously and try to identify some underlying diagrammatic mechanism. 
Here, the above QCD cancellation identities serve as a guiding principle.

\subsubsection{First observation: comparison of the propagators}
First, a propagator identity is constructed by following the identity \eqref{eq:canPropYM}. Indeed, 
we find
\begin{align}
\label{eq:canProp}
 \FGggPropContrCan{.5} & = \FGhhPropContrCan{.5}.
\end{align}
Here, the diagrammatics require some explanation. Whereas the gluon propagator has a single 
Lorentz index and the black triangle represents the longitudinal contraction of this index with the 
in-going momentum, the graviton propagator has two indices. The black triangle is meant to contract 
the two indices, say $μ$ and $ν$, and possesses an out-going ghost line that has a single index, say 
$λ$. In these conventions the black triangle represents $\nicefrac{(δ_μ^λp_ν + δ_ν^λp_μ)}{2}$, 
where $p$ denotes the in-going momentum. On the right hand side of the equation above, the dashed 
line 
represents the (graviton-) ghost propagator in straight analogy to \eqref{eq:canPropYM} for 
Yang--Mills. However, due to the more complicated numerator structure of the graviton propagator, 
the symmetrized longitudinal contraction gets modified by propagating through the propagator. The 
white triangle denotes this modified contraction and represents the expression 
$(δ_μ^λp_ν+δ_ν^λp_μ-η_{μν}p^λ)$. This needs to be carefully taken into account with dealing with 
Feynman diagrams with amputated external propagators.

\subsubsection{Enforcing the philosophy: the 3-graviton identity}

Since the graviton propagator allows for a propagation of the symmetrized longitudinal contraction, 
it is natural to ask whether this propagation also aligns with the graviton vertices. Indeed, by 
implementing a cancellation identity for each of the graviton vertices, relations between 
different diagrams and their divergences are established.

Hence, the next step is to derive a cancellation identity beginning with the 3-valent graviton 
vertex. It turns out, that the corresponding cancellation identity of Yang--Mills 
\eqref{eq:canTetraYM} can been seen as the blueprint for quantum gravity. After replacement of the 
gluon by gravitons (and their ghosts respectively), the only remaining unknown is the auxiliary 
vertex. It can however be constructed such that the cancellation identity holds, again in straight 
analogy to Yang--Mills diagrammatics
\begin{align}
\label{eq:canTri}
  \FGgggContrCan{0.75} & = 
\FGauxVertexCanL{0.75}+\FGghostVertexCanL{0.75}+\FGauxVertexCanR{0.75}+\FGghostVertexCanR{0.75}.
\end{align}
Here, it is worth remarking that the left hand side is a 3-graviton vertex with propagators 
attached to it, 
the white triangle is in accordance with \eqref{eq:canProp}, and the slashed edges that consist of 
a graviton and a straight half-edge indicate a cancelled graviton propagator. The actual expression 
for the auxiliary vertex is lengthy and will be discussed in more detail in future work. 
Nonetheless, it is worth remarking that as a function of the kinematic data, it reads 
$V_{\textnormal{aux}}(p_1,μ_1,ν_1;p_2,μ_2,ν_2;p_3,μ_3)$, where the semicolons separate the momentum 
and Lorentz indices of the graviton, straight, and ghost edges in this order. Whereas 
$V_{\textnormal{aux}}$ inherits the symmetry under permutations of $μ_i$ and $ν_i$ for both the 
graviton and straight edge (i.e.\ for $i=1,2$), it is not symmetric under permutation of these 
edges.

\subsubsection{An assessment: the 4-graviton identity}
At this stage, higher order cancellation identities will provide further non-trivial conditions 
to the auxiliary vertex constructed above. The 4-valent identity \eqref{eq:canTetraYM} serves as a 
first test and it translates to
\begin{align}
\label{eq:canTetra}
  \FGfourGContrCan{.75} &= - \FGfourGauxCanI{.66} - \FGfourGauxCanH{.75} - \FGfourGauxCanX{.75},
\end{align}
where the auxiliary vertex is determined through \eqref{eq:canTri}.

\subsubsection{Exercise fine judgment: higher order identities}

In contrast to the preceding identities, the cancellation of \eqref{eq:canPentaYM} does not 
directly translate to gravity. However, its interpretation that Yang--Mills does not require a 
5-valent vertex is also not correct for gravity. Hence it is reasonable to insert the longitudinal 
contraction of the 5-valent vertex on the left hand side of the equation and it turns out that 
corrects the 
equation resulting in the identity
\begin{align}
\label{eq:canPenta}
   \FGfiveGContrCan{.75} &= -\FGfiveGauxCanA{.75} - \FGfiveGauxCanB{.75} - \FGfiveGauxCanC{.75} -  
\FGfiveGauxCanD{.75}.\\
\intertext{From this identity and \eqref{eq:canTetra}, it is easy to deduce the pattern for 
vertices of higher valence: the symmetrized longitudinal contraction of a graviton vertex of 
valence $(n+1)$ yields all terms obtained by attaching the auxiliary vertex to a graviton vertex 
of valence $n$. We explicitly checked this for the 6-valent vertex and verified}
 \FGsixGContrCan{.75} &= - \FGsixGauxCanA{.75} - \FGsixGauxCanB{.75} -\FGsixGauxCanC{.75}\notag\\ 
\label{eq:canHexa}
&\quad\quad\quad -\FGsixGauxCanD{.75} -\FGsixGauxCanE{.75}.
\intertext{Finally, we also verified that the ghost cancellation identity of Yang--Mills 
\eqref{eq:canGhostYM} translates to the gravitation identity}
0 &= \FGghostGCanA{.75} - \FGghostGCanB{.75} + \FGghostGCanC{.75} - \FGghostGCanD{.75},
\label{eq:canGhost}
\end{align}
where all propagators of loose half-edges are considered to be amputated (as indicated due to 
the white triangles). We close this section with a table that lists the terms to be cancelled when  
the right hand side of an identity is subtracted from its left hand side. Here it is worth 
remarking that we always amputate propagators of external (i.e.\ loose) half-edges to avoid a 
tremendous increase in the number of terms generated by {\sc FORM}.

\begin{center}
\begin{tabular}{ |c|r| }
  \hline
  identity & \#cancelled terms \\\hline
  \eqref{eq:canProp} &  35\\
  \eqref{eq:canTri} &  3,168\\
  \eqref{eq:canTetra} & 30,888\\
  \eqref{eq:canPenta} & 607,488\\
  \eqref{eq:canHexa} & 12,105,600\\
  \eqref{eq:canGhost} & 1,536\\  \hline
\end{tabular}
\end{center}
Note that over 12 million terms have to match up for \eqref{eq:canHexa} and checking explicitly that 
this is the case non-trivially substantiates our initial approach.

\section{Discussion}

The cancellation identities of Yang--Mills theory can be translated to identities for pure 
quantum gravity in the classical approach by Capper, Leibbrandt, and Medrano \cite{Capper:1973pv}. 
The reason for choosing this setting is due to its relatively simple gauge 
\eqref{eq:EHgaugeFixing} that induces only a single ghost-vertex. This allowed us to establish the 
identities and success of this approach will serve as a basis for future extensions. In close 
analogy to Yang--Mills, the cancellation identity for the 3-graviton vertex determines an auxiliary 
vertex \eqref{eq:canTri}. However, the generalization of the longitudinal contraction of a graviton 
propagator requires the introduction of an additional contraction, which we represented by a white 
triangle \eqref{eq:canProp}. We demonstrated that the cancellation identities for the 4-graviton as 
well as the ghost vertex exactly match their Yang--Mills analogues. Also higher order identities 
have been constructed by considering longitudinal contracted graviton vertices of that order (which 
do not have analogues in Yang--Mills). Notably, the cancellation identities 
(\ref{eq:canTetra}--\ref{eq:canHexa}) demonstrate that a gravitational vertex of a certain valence 
relates to vertices of lower valence. This observation parallels to Yang--Mills theory, where the 
relation for 4-valent vertex \eqref{eq:canTetraYM} can be understood as strong evidence to the 
existence of a corolla differential, that produces 4-valent vertices from purely 3-valent graphs. 
Our findings support the conjecture that a full generation of gravitational amplitudes from 
3-valent graphs by an appropriate corolla differential is possible.

Intuitively, it might come as a surprise that exactly those diagrammatic identities which 
guarantee the renormalizability of Yang--Mills theory do generalize to gravity. However, the 
existence of diagrammatic cancellations seems plausible in the light of \cite{DeWitt:1967ub} and the 
presence of a BRST invariance and  Ward identities 
\cite{Capper:1974vb,Delbourgo:1976xd,Delbourgo:1985wz}. We do consider the exact mathematical 
connection of diagrammatics and the BRST cohomology an exciting topic for further investigation. 
The cancellation identities derived here open up the possibility to study the gauge dependence by 
diagrammatic techniques \cite{Kissler:2016tne,Kissler:2018lnn}, where it was also empirically 
observed that this mechanism applies to more general situations 
\cite{Zerf:2018csr,Ihrig:2019kfv,Gracey:2020kbb}. For gravity, this has potential to determine the 
divergences that are of genuine physical significance or to diagrammatically construct generalized 
Slavnov--Taylor identities for gravity and to study their self-consistency off-shell as was recently 
accomplished for Yang--Mills \cite{Gracey:2019mix}. 

Another significance of the demonstrated cancellations stems from the diagrams with cancelled 
propagators. In the sense of graph theory, these define an operation on graphs by edge-collapsing.  
In combination with the ghost identity, this informs the definition of a gravitational graph and 
cycle homology as in the Yang--Mills case \cite{Kreimer:2012jw,Berghoff2020} and possibly enables 
the construction of a corolla polynomial for gravity as extension of 
\cite{Kreimer:2012eh,Kreimer:2012jw,Prinz:2016fka,Kreimer:2018lyh}. For this, it worth stressing 
that on the graphical level, the identity for the gravitational ghosts \eqref{eq:canGhost} matches 
exactly its analogue \eqref{eq:canGhostYM} of Yang--Mills theory.

In the future, our approach might benefit from the vast range of double-copy techniques 
\cite{Bern:2008qj,Bern:2010ue,Bern:2010yg,Bern:2019prr}, which excitingly have been employed in the 
off-shell regime in \cite{Anastasiou:2018rdx,Borsten:2020xbt}, to mention but a few. So far, it has 
been our objectives to consider a Lagrangian as close as possible to Einstein--Hilbert theory and 
concentrate on a simple gauge-fixing, which minimizes the number of graviton-ghost interactions for 
the sake of simple cancellation identities. Within this class of gauges, we have been studying the 
most general parametrization in order to allow the gauge parameter $ξ$ to absorb off-shell 
singularities that inhere in loop diagrams. In this tensor-density-based expansion, the observed 
diagrammatic relations encourage interpretation of a gravitational vertex as a certain convolution 
of gluonic kinematics. However, from the analytic form of the graviton propagator as displayed 
in \cite{Capper:1978yf}, it seems reasonable to expect such an endeavour to require the introduction 
of supplementary fields or even a change of gauge.

\section*{Acknowledgments}
My understanding of gravity was greatly enhanced thanks to discussions with D.\ Prinz. Further 
thanks to P.\ Balduf for useful discussions on amputated Green's functions, D.\ Kreimer for 
encouragement, and J.A.\ Gracey for useful comments on the manuscript. The use of axodraw2 
\cite{Collins:2016aya} is acknowledged. This work was carried out with the support of Deutsche  
Forschungsgemeinschaft  (DFG Grant KR1401/5-2).

\bibliographystyle{elsarticle-num}
\bibliography{gravity}

\begin{thebibliography}{10}
\expandafter\ifx\csname url\endcsname\relax
  \def\url#1{\texttt{#1}}\fi
\expandafter\ifx\csname urlprefix\endcsname\relax\def\urlprefix{URL }\fi
\expandafter\ifx\csname href\endcsname\relax
  \def\href#1#2{#2} \def\path#1{#1}\fi

\bibitem{Bloch:2005bh}
S.~Bloch, H.~Esnault, D.~Kreimer, {On Motives associated to graph polynomials},
  Commun. Math. Phys. 267 (2006) 181--225.
\newblock \href {http://arxiv.org/abs/math/0510011}
  {\path{arXiv:math/0510011}}, \href
  {http://dx.doi.org/10.1007/s00220-006-0040-2}
  {\path{doi:10.1007/s00220-006-0040-2}}.

\bibitem{Panzer:2015ida}
E.~Panzer, {Feynman integrals and hyperlogarithms}, Ph.D. thesis, Humboldt U.
  (2015).
\newblock \href {http://arxiv.org/abs/1506.07243} {\path{arXiv:1506.07243}},
  \href {http://dx.doi.org/10.18452/17157} {\path{doi:10.18452/17157}}.

\bibitem{Bloch:2015efx}
S.~Bloch, D.~Kreimer, {Cutkosky Rules and Outer Space}\href
  {http://arxiv.org/abs/1512.01705} {\path{arXiv:1512.01705}}.

\bibitem{Brown:2015fyf}
F.~Brown, {Feynman amplitudes, coaction principle, and cosmic Galois group},
  Commun. Num. Theor. Phys. 11 (2017) 453--556.
\newblock \href {http://arxiv.org/abs/1512.06409} {\path{arXiv:1512.06409}},
  \href {http://dx.doi.org/10.4310/CNTP.2017.v11.n3.a1}
  {\path{doi:10.4310/CNTP.2017.v11.n3.a1}}.

\bibitem{Schnetz:2017bko}
O.~Schnetz, {The Galois coaction on the electron anomalous magnetic moment},
  Commun. Num. Theor. Phys. 12 (2018) 335--354.
\newblock \href {http://arxiv.org/abs/1711.05118} {\path{arXiv:1711.05118}},
  \href {http://dx.doi.org/10.4310/CNTP.2018.v12.n2.a4}
  {\path{doi:10.4310/CNTP.2018.v12.n2.a4}}.

\bibitem{Schnetz:2019cab}
O.~Schnetz, {Geometries in perturbative quantum field theory}\href
  {http://arxiv.org/abs/1905.08083} {\path{arXiv:1905.08083}}.

\bibitem{Panzer:2019yxl}
E.~Panzer, {Hepp's bound for Feynman graphs and matroids}\href
  {http://arxiv.org/abs/1908.09820} {\path{arXiv:1908.09820}}.

\bibitem{Klausen:2019hrg}
R.~P. Klausen, {Hypergeometric Series Representations of Feynman Integrals by
  GKZ Hypergeometric Systems}, JHEP 04 (2020) 121.
\newblock \href {http://arxiv.org/abs/1910.08651} {\path{arXiv:1910.08651}},
  \href {http://dx.doi.org/10.1007/JHEP04(2020)121}
  {\path{doi:10.1007/JHEP04(2020)121}}.

\bibitem{Tapuskovic:2019cpr}
M.~Tapušković, {Motivic Galois coaction and one-loop Feynman graphs}\href
  {http://arxiv.org/abs/1911.01540} {\path{arXiv:1911.01540}}.

\bibitem{Kreimer:2012jw}
D.~Kreimer, M.~Sars, W.~van Suijlekom, {Quantization of gauge fields, graph
  polynomials and graph homology}, Annals Phys. 336 (2013) 180--222.
\newblock \href {http://arxiv.org/abs/1208.6477} {\path{arXiv:1208.6477}},
  \href {http://dx.doi.org/10.1016/j.aop.2013.04.019}
  {\path{doi:10.1016/j.aop.2013.04.019}}.

\bibitem{Berghoff2020}
M.~Berghoff, A.~Knispel, {Complexes of marked graphs in gauge theory}, Letters
  in Mathematical Physics\href {http://arxiv.org/abs/1908.06640}
  {\path{arXiv:1908.06640}}, \href
  {http://dx.doi.org/10.1007/s11005-020-01301-0}
  {\path{doi:10.1007/s11005-020-01301-0}}.

\bibitem{Prinz:2016fka}
D.~Prinz, {The Corolla Polynomial for spontaneously broken Gauge Theories},
  Math. Phys. Anal. Geom. 19~(3) (2016) 18.
\newblock \href {http://arxiv.org/abs/1603.03321} {\path{arXiv:1603.03321}},
  \href {http://dx.doi.org/10.1007/s11040-016-9222-0}
  {\path{doi:10.1007/s11040-016-9222-0}}.

\bibitem{Cheung:2017kzx}
C.~Cheung, G.~N. Remmen, {Hidden Simplicity of the Gravity Action}, JHEP 09
  (2017) 002.
\newblock \href {http://arxiv.org/abs/1705.00626} {\path{arXiv:1705.00626}},
  \href {http://dx.doi.org/10.1007/JHEP09(2017)002}
  {\path{doi:10.1007/JHEP09(2017)002}}.

\bibitem{Prinz:2018dhq}
D.~Prinz, {Algebraic Structures in the Coupling of Gravity to Gauge Theorie
  s}\href {http://arxiv.org/abs/1812.09919} {\path{arXiv:1812.09919}}.

\bibitem{Prinz:2020nru}
D.~Prinz, {Gravity-Matter Feynman Rules for any Valence}\href
  {http://arxiv.org/abs/2004.09543} {\path{arXiv:2004.09543}}.

\bibitem{Capper:1973pv}
D.~Capper, G.~Leibbrandt, M.~Ramon~Medrano, {Calculation of the graviton
  selfenergy using dimensional regular ization}, Phys. Rev. D 8 (1973)
  4320--4331.
\newblock \href {http://dx.doi.org/10.1103/PhysRevD.8.4320}
  {\path{doi:10.1103/PhysRevD.8.4320}}.

\bibitem{Pascual:1984zb}
P.~Pascual, R.~Tarrach, {QCD: Renormalization for the Practitioner}, Lect.
  Notes Phys. 194 (1984) 1--277.

\bibitem{Kreimer:2005rw}
D.~Kreimer, {Anatomy of a gauge theory}, Annals Phys. 321 (2006) 2757--2781.
\newblock \href {http://arxiv.org/abs/hep-th/0509135}
  {\path{arXiv:hep-th/0509135}}, \href
  {http://dx.doi.org/10.1016/j.aop.2006.01.004}
  {\path{doi:10.1016/j.aop.2006.01.004}}.

\bibitem{Slavnov:1972fg}
A.~Slavnov, {Ward Identities in Gauge Theories}, Theor. Math. Phys. 10 (1972)
  99--107.
\newblock \href {http://dx.doi.org/10.1007/BF01090719}
  {\path{doi:10.1007/BF01090719}}.

\bibitem{Taylor:1971ff}
J.~C. Taylor, {Ward Identities and Charge Renormalization of the Yang-Mills
  Field}, Nucl. Phys. B 33 (1971) 436--444.
\newblock \href {http://dx.doi.org/10.1016/0550-3213(71)90297-5}
  {\path{doi:10.1016/0550-3213(71)90297-5}}.

\bibitem{Gracey:2019mix}
J.~A. Gracey, H.~Kißler, D.~Kreimer, {Self-consistency of off-shell
  Slavnov-Taylor identities in QCD}, Phys. Rev. D 100~(8) (2019) 085001.
\newblock \href {http://arxiv.org/abs/1906.07996} {\path{arXiv:1906.07996}},
  \href {http://dx.doi.org/10.1103/PhysRevD.100.085001}
  {\path{doi:10.1103/PhysRevD.100.085001}}.

\bibitem{Lautrup:1977ty}
B.~Lautrup, {Of Ghoulies and Ghosties: An Introduction to QCD}.

\bibitem{Cvitanovic:1977qu}
P.~Cvitanović, {Quantum Chromodynamics on the Mass Shell}, Nucl. Phys. B 130
  (1977) 114--144.
\newblock \href {http://dx.doi.org/10.1016/0550-3213(77)90396-0}
  {\path{doi:10.1016/0550-3213(77)90396-0}}.

\bibitem{tHooft:1974toh}
G.~'t~Hooft, M.~Veltman, {One loop divergencies in the theory of gravitation},
  Ann. Inst. H. Poincare 20 (1974) 69--94.

\bibitem{Goroff:1985sz}
M.~H. Goroff, A.~Sagnotti, {Quantum Gravity at Two Loops}, Phys. Lett. B 160
  (1985) 81--86.
\newblock \href {http://dx.doi.org/10.1016/0370-2693(85)91470-4}
  {\path{doi:10.1016/0370-2693(85)91470-4}}.

\bibitem{Goroff:1985th}
M.~H. Goroff, A.~Sagnotti, {The Ultraviolet Behavior of Einstein Gravity},
  Nucl. Phys. B 266 (1986) 709--736.
\newblock \href {http://dx.doi.org/10.1016/0550-3213(86)90193-8}
  {\path{doi:10.1016/0550-3213(86)90193-8}}.

\bibitem{vandeVen:1991gw}
A.~van~de Ven, {Two loop quantum gravity}, Nucl. Phys. B 378 (1992) 309--366.
\newblock \href {http://dx.doi.org/10.1016/0550-3213(92)90011-Y}
  {\path{doi:10.1016/0550-3213(92)90011-Y}}.

\bibitem{Weinberg:2009bg}
S.~Weinberg, {Effective Field Theory, Past and Future}, PoS CD09 (2009) 001.
\newblock \href {http://arxiv.org/abs/0908.1964} {\path{arXiv:0908.1964}},
  \href {http://dx.doi.org/10.22323/1.086.0001}
  {\path{doi:10.22323/1.086.0001}}.

\bibitem{Weinberg:2016kyd}
S.~Weinberg, {Effective field theory, past and future}, Int. J. Mod. Phys. A
  31~(06) (2016) 1630007.
\newblock \href {http://dx.doi.org/10.1142/S0217751X16300076}
  {\path{doi:10.1142/S0217751X16300076}}.

\bibitem{Landau:1951dk}
L.~D. Landau, E.~M. Lifshitz, The Classical Theory of Fields, Pergamon Press,
  Oxford, 1951.

\bibitem{Capper:1978yf}
D.~Capper, M.~Namazie, {A General Gauge Calculation of the Graviton
  Selfenergy}, Nucl. Phys. B 142 (1978) 535--547.
\newblock \href {http://dx.doi.org/10.1016/0550-3213(78)90229-8}
  {\path{doi:10.1016/0550-3213(78)90229-8}}.

\bibitem{Vermaseren:2000nd}
J.~A.~M. Vermaseren, {New features of FORM}\href
  {http://arxiv.org/abs/math-ph/0010025} {\path{arXiv:math-ph/0010025}}.

\bibitem{Ruijl:2017dtg}
B.~Ruijl, T.~Ueda, J.~A.~M. Vermaseren, {FORM version 4.2}\href
  {http://arxiv.org/abs/1707.06453} {\path{arXiv:1707.06453}}.

\bibitem{Kreimer:2008dg}
D.~Kreimer, {Not so non-renormalizable gravity}, in: {Conference on Recent
  Developments in Quantum Field Theory}, 2008, pp. 155--162.
\newblock \href {http://arxiv.org/abs/0805.4545} {\path{arXiv:0805.4545}},
  \href {http://dx.doi.org/10.1007/978-3-7643-8736-5\_9}
  {\path{doi:10.1007/978-3-7643-8736-5\_9}}.

\bibitem{DeWitt:1967ub}
B.~S. DeWitt, {Quantum Theory of Gravity. 2. The Manifestly Covariant Theory},
  Phys. Rev. 162 (1967) 1195--1239.
\newblock \href {http://dx.doi.org/10.1103/PhysRev.162.1195}
  {\path{doi:10.1103/PhysRev.162.1195}}.

\bibitem{Capper:1974vb}
D.~Capper, M.~Medrano, {Gravitational slavnov-ward identities}, Phys. Rev. D 9
  (1974) 1641--1647.
\newblock \href {http://dx.doi.org/10.1103/PhysRevD.9.1641}
  {\path{doi:10.1103/PhysRevD.9.1641}}.

\bibitem{Delbourgo:1976xd}
R.~Delbourgo, M.~Ramon~Medrano, {Becchi-Rouet-Stora Gauge Identities for
  Gravity}, Nucl. Phys. B 110 (1976) 467--472.
\newblock \href {http://dx.doi.org/10.1016/0550-3213(76)90235-2}
  {\path{doi:10.1016/0550-3213(76)90235-2}}.

\bibitem{Delbourgo:1985wz}
R.~Delbourgo, T.~Matsuki, {Gravitational counterterms and Becchi-Rouet-Stora
  symmetry}, Phys. Rev. D 32 (1985) 2579.
\newblock \href {http://dx.doi.org/10.1103/PhysRevD.32.2579}
  {\path{doi:10.1103/PhysRevD.32.2579}}.

\bibitem{Kissler:2016tne}
H.~Kißler, D.~Kreimer, {Diagrammatic Cancellations and the Gauge Dependence of
  QED}, Phys. Lett. B 764 (2017) 318--321.
\newblock \href {http://arxiv.org/abs/1607.05729} {\path{arXiv:1607.05729}},
  \href {http://dx.doi.org/10.1016/j.physletb.2016.11.052}
  {\path{doi:10.1016/j.physletb.2016.11.052}}.

\bibitem{Kissler:2018lnn}
H.~Kißler, {On the gauge dependence of Quantum Electrodynamics}, PoS LL2018
  (2018) 032.
\newblock \href {http://arxiv.org/abs/1808.00291} {\path{arXiv:1808.00291}},
  \href {http://dx.doi.org/10.22323/1.303.0032}
  {\path{doi:10.22323/1.303.0032}}.

\bibitem{Zerf:2018csr}
N.~Zerf, P.~Marquard, R.~Boyack, J.~Maciejko, {Critical behavior of the
  QED$_3$-Gross-Neveu-Yukawa model at four loops}, Phys. Rev. B 98~(16) (2018)
  165125.
\newblock \href {http://arxiv.org/abs/1808.00549} {\path{arXiv:1808.00549}},
  \href {http://dx.doi.org/10.1103/PhysRevB.98.165125}
  {\path{doi:10.1103/PhysRevB.98.165125}}.

\bibitem{Ihrig:2019kfv}
B.~Ihrig, N.~Zerf, P.~Marquard, I.~F. Herbut, M.~M. Scherer, {Abelian Higgs
  model at four loops, fixed-point collision and deconfined criticality}, Phys.
  Rev. B 100~(13) (2019) 134507.
\newblock \href {http://arxiv.org/abs/1907.08140} {\path{arXiv:1907.08140}},
  \href {http://dx.doi.org/10.1103/PhysRevB.100.134507}
  {\path{doi:10.1103/PhysRevB.100.134507}}.

\bibitem{Gracey:2020kbb}
J.~A. Gracey, {Six dimensional ultraviolet completion of the $CP(N)$ $\sigma$
  model at two loops}\href {http://arxiv.org/abs/2003.06618}
  {\path{arXiv:2003.06618}}.

\bibitem{Kreimer:2012eh}
D.~Kreimer, K.~Yeats, {Properties of the corolla polynomial of a 3-regular
  graph}\href {http://arxiv.org/abs/1207.5460} {\path{arXiv:1207.5460}}.

\bibitem{Kreimer:2018lyh}
D.~Kreimer, {The corolla polynomial: a graph polynomial on half-edges}, PoS
  LL2018 (2018) 068.
\newblock \href {http://arxiv.org/abs/1807.02385} {\path{arXiv:1807.02385}},
  \href {http://dx.doi.org/10.22323/1.303.0068}
  {\path{doi:10.22323/1.303.0068}}.

\bibitem{Bern:2008qj}
Z.~Bern, J.~J.~M. Carrasco, H.~Johansson, {New Relations for Gauge-Theory
  Amplitudes}, Phys. Rev. D 78 (2008) 085011.
\newblock \href {http://arxiv.org/abs/0805.3993} {\path{arXiv:0805.3993}},
  \href {http://dx.doi.org/10.1103/PhysRevD.78.085011}
  {\path{doi:10.1103/PhysRevD.78.085011}}.

\bibitem{Bern:2010ue}
Z.~Bern, J.~J.~M. Carrasco, H.~Johansson, {Perturbative Quantum Gravity as a
  Double Copy of Gauge Theory}, Phys. Rev. Lett. 105 (2010) 061602.
\newblock \href {http://arxiv.org/abs/1004.0476} {\path{arXiv:1004.0476}},
  \href {http://dx.doi.org/10.1103/PhysRevLett.105.061602}
  {\path{doi:10.1103/PhysRevLett.105.061602}}.

\bibitem{Bern:2010yg}
Z.~Bern, T.~Dennen, Y.-t. Huang, M.~Kiermaier, {Gravity as the Square of Gauge
  Theory}, Phys. Rev. D 82 (2010) 065003.
\newblock \href {http://arxiv.org/abs/1004.0693} {\path{arXiv:1004.0693}},
  \href {http://dx.doi.org/10.1103/PhysRevD.82.065003}
  {\path{doi:10.1103/PhysRevD.82.065003}}.

\bibitem{Bern:2019prr}
Z.~Bern, J.~J. Carrasco, M.~Chiodaroli, H.~Johansson, R.~Roiban, {The Duality
  Between Color and Kinematics and its Applications}\href
  {http://arxiv.org/abs/1909.01358} {\path{arXiv:1909.01358}}.

\bibitem{Anastasiou:2018rdx}
A.~Anastasiou, L.~Borsten, M.~J. Duff, S.~Nagy, M.~Zoccali, {Gravity as Gauge
  Theory Squared: A Ghost Story}, Phys. Rev. Lett. 121~(21) (2018) 211601.
\newblock \href {http://arxiv.org/abs/1807.02486} {\path{arXiv:1807.02486}},
  \href {http://dx.doi.org/10.1103/PhysRevLett.121.211601}
  {\path{doi:10.1103/PhysRevLett.121.211601}}.

\bibitem{Borsten:2020xbt}
L.~Borsten, S.~Nagy, {The pure BRST Einstein-Hilbert Lagrangian from the
  double-copy to cubic order}, JHEP 07 (2020) 093.
\newblock \href {http://arxiv.org/abs/2004.14945} {\path{arXiv:2004.14945}},
  \href {http://dx.doi.org/10.1007/JHEP07(2020)093}
  {\path{doi:10.1007/JHEP07(2020)093}}.

\bibitem{Collins:2016aya}
J.~C. Collins, J.~A.~M. Vermaseren, {Axodraw Version 2}\href
  {http://arxiv.org/abs/1606.01177} {\path{arXiv:1606.01177}}.

\end{thebibliography}

\end{document}